# Design of a 10 MeV normal conducting CW proton linac based on equidistant multi-gap CH cavities[*]

LI Zhi-Hui (李智慧)

[1)] Institute of Nuclear Science and Technology, Sichuan University, Chengdu, 610065, China

Abstract: The continue wave (CW) high current proton linac has wide applications as the front end of the high power proton machines. The low energy part is the most difficult one and there is no widely accepted solution yet. Based on the analysis of the focusing properties of the CW low energy proton linac, a 10 MeV low energy normal conducting proton linac based on equidistant seven-gap Cross-bar H-type (CH) cavities is proposed. The linac is composed of ten 7-gap CH cavities and the transverse focusing is maintained by the quadrupole doublets located between cavities. The total length of the linac is less than 6 meters and the average acceleration gradient is about 1.2 MeV/m. The electromagnetic properties of the cavities are investigated by Microwave Studio. At the nominal acceleration gradient the maximum surface electric field in the cavities is less than 1.3 times Kilpatrick limit, and the Ohmic loss of each cavity is less than 35 kW. The multi-particle beam dynamics simulations are performed with the help of the Tracewin code, the results show that the beam dynamics of the linac is quite stable, and the linac has the capability to accelerate up to 30 mA beam with acceptable dynamics behavior.

## 1 Introduction

The rising concern about the greenhouse effect and the limited conservations of the fossil fuels pleads for a severe reduction in the use of fossil fuels [1]. One possible way to allow the production of energy essential to the development of developing countries, which account for the majority of humankind, without catastrophically increasing greenhouse gas emissions might be to rely increasingly on nuclear fission energy based on its property of negligible contribution to this effect [2]. China has increased its investment on nuclear power in the past two decades and this trend is foresee to be continued in the following several decades. However the nuclear electric plant based on light-water reactors will produce nuclear wastes, which will need more than 1 million years to decay to the reference radio toxicity level of the Uranium ore [3]. Thus it is more and more urgent to find a consensual solution to transmute the long-lived radioactive waste in order to keep the nuclear energy sustainable. An Accelerator Driver System (ADS), coupling a proton accelerator, a spallation target and a sub-critical reactor, is a very promising technology for nuclear waste transmutation [4].

The proton beams delivered to the spallation target should have energy more than 600MeV in order to optimize the number of neutrons produced per MeV of incident energy; the total beam power should exceed 10MW, possible more and with high beam availability. It is clear that the requirements for both beam power and beam availability are all beyond the capability of existing accelerators. The development of the superconducting techniques, especially the middle beta cavities makes it possible to build a superconducting linear accelerator from several tens of MeV [5, 6, 7]. Although the high frequency (>300MHz) low beta structures, motivated by the acceleration of high intensity proton beams, have attracted interests in Argonne National Laboratory, Los Alamos National Laboratory and other accelerator labs since 1990's [8], and China has achieved great progress in low beta superconducting structures since 2010, the integrated properties of low filling ratio makes the acceleration gradient of low energy superconducting linac has no obvious advantage compared with the normal conducting one and the beam current is limited to only about 20-30 mA [9]. A study program on ADS organized by Chinese Academy of Sciences (CAS) was proposed and participated by several institutes of CAS, which is concentrated on designing and developing a total superconducting 15 MW continue wave (CW) proton linac except the radio frequency quadruple (RFQ) accelerators. At the same time, under the support of the National Natural Science Foundation of China, the exploration of the possibility to design and built a low energy CW normal conducting proton linac with energy between 10 to 20 MeV as the front end of the ADS driver linac is also under way. In this paper, a scheme based on equidistant multi-gap Cross-bar H-type (CH) cavities is presented.

---
[*] Supported by National Natural Science Foundation of China (11375122, 91126003)
1) E-mail: lizhihui@scu.edu.cn

## 2 Why equidistant multi-gap CH cavities?

Although the acceleration gradient of CW machine is quite low, usually just around 1 MV/m, only 1/3 or 1/4 that of the pulsed machine, and the peak power is decreased dramatically so that it can be properly cooled down, it still hard to image to build a long drift tube linac (DTL) tank like the pulsed one. It is partly because that the very large average power gives a lot of challenges in design of the RF system, such as amplifier, power coupler and so on. The long tank with more than 20 gaps will also decrease the RF stability of the cavity since the very small gaps between neighboring modes. Furthermore, the decreased accelerating gradient not only decreases the dissipated power, but also decreases the longitudinal focusing and transverse defocusing strength of the RF field. As a result, it is possible to have longer transverse focusing period than the pulsed one, it is natural to apply a series of cavities with multi-gaps and without transverse focusing elements in them, the transverse focusing can be maintained by the magnetic quadrupole lenses between cavities, and the number of gaps within each cavity can be determined by the longitudinal and transverse focusing requirements. By this way, on one side the longitudinal focusing and transverse focusing would be more balance and which is very important for high current beams. We have simulated the behavior of a 3 MeV, 20 mA proton beam with 3σ truncated Gaussian distribution transporting through a period focusing channel. The focusing channel is consisted of 74 cells. Each cell is composed in a magnetic quadrupole doublet and a RF gap. During simulation we set the RF phase of the RF gaps at -90 degree, so that they just work as bunchers and without acceleration on the particles. The widely used Tracewin code [10] with 2D Particle-In-Cell (PIC) space charge routine is used for simulation. Four different cases are simulated: Case-I, the transverse and longitudinal phase advance per period is 67 and 75 degree; Case-II, the transverse and longitudinal is 24 and 27 degree, the ratio of the transverse and longitudinal focusing strength is the same as Case-I, only the focusing strength is decreased; Case-III and Case-IV belong to the unbalanced focusing cases and the transverse and longitudinal phase advances per period are 24 and 75 for Case-III and 67 and 27 for Case-IV. The normalized rms emittance evolutions for the for cases are shown in Fig.1 and reveal that if the transverse and longitudinal focusing is balanced (Case-I and Case-II), the emittances growth are well controlled (less than 10%) in both transverse and longitudinal plans, even for the very weak focusing Case-II; if the transverse and longitudinal focusing is unbalanced (Case-III and Case-IV), the emittance of the weak focusing plan increases significantly (transverse of Case-III and longitudinal of Case-IV). Further study shows the nonlinear space charge force plays important roles in the emittance growth [9]. On the other side, since there is no transverse focusing elements in the drift tubes, the radial dimension of the drift tubes can be decreased and slim drift tubes can be used, which will increase the effective shunt impedance greatly. Because the cavity only contains several gaps (usually less than 10), so the total power per cavity is also limited. This will also decrease the difficulties in the design and fabrication of the power coupler.

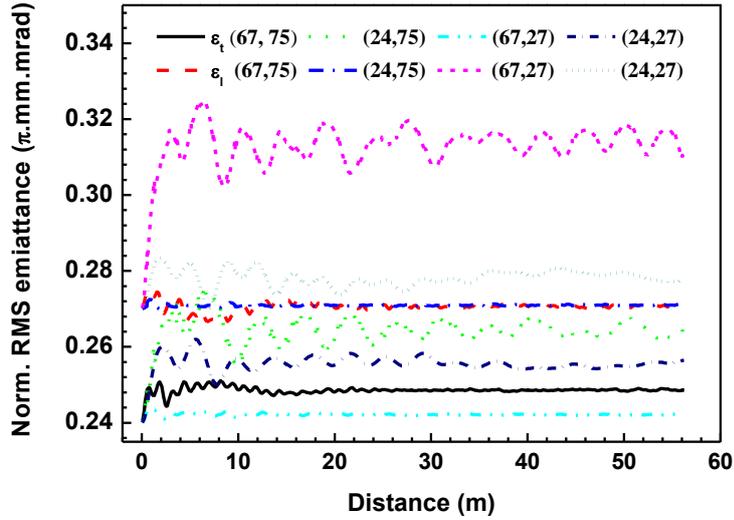

Fig.1 the evolution of normalized RMS emittances of a 20 mA proton beam transport through a focusing channel.

For CW normal conducting linacs, one of the biggest challenges comes from how to decrease the dissipated power density on the cavity surface, so that after proper design of the cooling channel, the power can be removed out of the cavity. The CH structure matches this requirement perfectly since its high effective shunt impedance compared with the widely used Alvarez type drift tube linac structure [11]. Even compared with multi-tank DTL (MTDTL) structures, which is characterized of short DTL tank with 4 to 10 gaps in it, and the drift tubes are free of transverse focusing elements, so that the effective shunt impedance is much higher than that of the traditional DTL one, the effective shunt impedance of CH structure is still more than two times of it. Of cause what is really important concerned with cooling is the power density, and the effective shunt impedance just gives us some idea about the total power dissipation. By adoption of slim drift tubes, the capacitance of acceleration gaps is decreased, so that the MTDTL has larger transverse dimension compared with that of the CH structure. For example as Table 1 shows, the radius of a 325MHz MTDTL cavity with geometry beta 0.087 is around 340 mm, while for a CH structure with same frequency and geometry beta, the cavity radius is only about 161 mm, but our simulation results show that not only the total dissipated power of CH structure is less than that of the MTDTL one, but also the maximum surface electric and magnetic field are smaller than that of the MTDTL structure. It means that the power is more uniformly distributed within CH structure and this makes it easier to be cooled down. By adoption equidistant structure, the geometries of drift tubes and gaps within one cavity are identical, and the field distribution can be adjusted by the two end cells. This not only eases the fabrication and installation of the drift tubes, but also decreases the number of the movable devices for field tuning, and it would be very helpful for stable working of a cavity in CW mode. Based on the reasons mentioned above, it is natural to choose the equidistant multi-gap CH cavities as the acceleration structure for a low energy high current CW normal conducting proton linac.

Table 1: comparison of the main parameters of MTDTL and CH structures

| Parameters | DTL | CH |
|---|---|---|
| Frequency (MHz) | 325.0 | 325.0 |
| Cavity radius (mm) | 340.0 | 161.0 |
| Aperture radius (mm) | 10 | 10 |
| Cell length (mm) | 80 | 40 |
| Cell Numbers | 5 | 5 |
| Gap length (mm) | 18 | 18 |
| 1 MV/m gradient 1m structure | | |
| Power (kW/m) | 35.02 | 15.96 |
| Max surface electric field (MV/m) | 13.46 | 8.35 |
| Max surface magnetic field (A/m) | 4610.73 | 4269.76 |

# 3 Cavity properties and lattice layout

The first parameter need to be decided before the cavity design is performed is how many cells each cavity should have? This number is decided by the expected acceleration gradient of the linac. Since this machine is a high current proton one, the requirements for stable beam dynamics have to be satisfied. One of the most important requirements for stable beam dynamics is that the zero current phase advances per period in three directions should be kept below 90 degrees. In our case, the focusing period is composed of one cavity and the quadruples between cavities. If we want to achieve an effective average acceleration gradient of about 1 MV/m, which is comparable with that obtained by the low beta superconducting linac for China ADS injectors under development [12], then we can determine the number of cells per cavity. We found under this condition, the seven cell cavity is perfect and the corresponding maximum surface electric field is less than 1.3 times Kilpatrick filed, which is thought to be a safe number in CW RFQ design. Fig. 2 shows the cavity geometry with cell length 43 mm, the corresponding geometry beta is 0.093. The cavity is composed of a cylinder as the vacuum chamber, which is loaded by the alternatively oriented in x and y directions bars, the bars are all identical and are composed of a cylinder which is connected to the outer cylinder cavity on one side and the other side is connected with a cone, the cone is smoothly connected to another cylinder with small radius, which is connected to the drift tube on the other side. The drift tubes are located at the axis of the cavity. On the two sides of the cavity, two enlarged cylinders with radius of 80 mm are intruded into the cavity with length of 120 to 130 mm to obtain an approximately uniform longitudinal field distribution along axis, just as Fig. 2 shows. There are ten cavities applied to accelerate the beam from 3 MeV to 10 MeV, and only the period length and the gap width are changed in different cavities. The main parameters of the cavities are shown in table 2. The period length are almost linearly increased from 37 mm to 65 mm and the minimum drift tube length in first cavity is 20 mm, which is equal to the inner drift tube diameter to make sure the field in the drift tubes can be well shielded. The field levels in each cavity are determined by the requirements of the longitudinal phase advances, that means the maximum zero current phase advance per period is less than 90 degree and the phase advance per meter is smoothly changed.

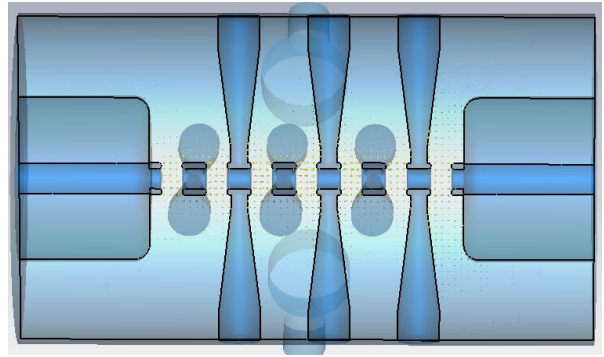

Fig. 2 the geometry of the CH cavity

Table 1 the parameters of the cavities

| Cell length (mm) | Diameter (mm) | Cavity Length(mm) | Eff. voltage (MV) | Acce. Grad. (MeV/m) | $K_p$ | Power (kW) |
|---|---|---|---|---|---|---|
| 37 | 301 | 490 | 0.56 | 0.93 | 1.1 | 18.3 |
| 40 | 312 | 510 | 0.68 | 1.12 | 1.2 | 24.3 |
| 43 | 320 | 530 | 0.76 | 1.24 | 1.3 | 27.3 |
| 46 | 326 | 550 | 0.80 | 1.15 | 1.3 | 25.8 |
| 50 | 324 | 570 | 0.81 | 1.22 | 1.3 | 26.2 |
| 53 | 333 | 590 | 0.84 | 1.24 | 1.3 | 26.8 |
| 56 | 335 | 610 | 0.86 | 1.22 | 1.3 | 26.7 |
| 59 | 340 | 620 | 0.90 | 1.26 | 1.2 | 29.2 |
| 62 | 341 | 640 | 0.94 | 1.27 | 1.2 | 31.0 |
| 65 | 344 | 670 | 0.99 | 1.28 | 1.3 | 33.7 |

Notes: $K_p$: Kilpatrick factor and the acceleration gradient is calculated with the cavity total length

The structure of the focusing lattice is straight forward; the focusing period is composed of one cavity and two quadruples as Fig. 3 shows. The cavity is set to work at synchronous phase from -35 degree at the first cavity to smoothly increase to -30 degree at the last cavity so that the beam can be properly focused in longitudinal, and the transverse focusing is performed by the quadrupole doublet. Each quadrupole lens has an effective length of 80 mm and separated with 20 mm drift space between them. The maximum gradient of the quadrupole lens is about 35 T/m. The zero current phase advances of longitudinal and transverse directions are shown in Fig. 4. With such settings, the total length of the linac is about 5.8 meters and the average acceleration gradient is about 1.2 MeV/m.

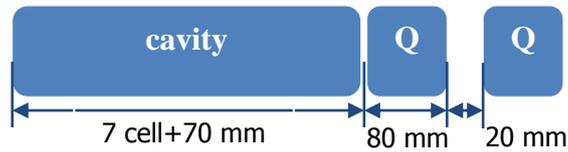

Fig.3 the layout of the focusing structure

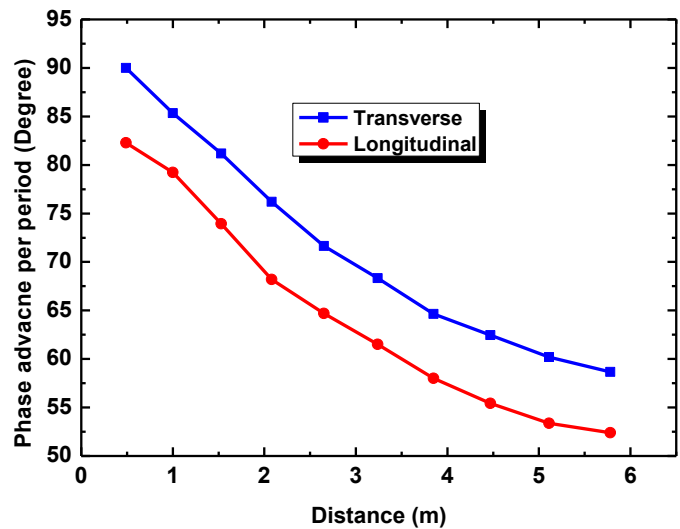

Fig.4 phase advance per period along the linac

## 4 Beam dynamics simulations results and discussion

In order to check the dynamic stability and the current limits, the multi-particles simulations are performed with Tracewin [10], which is one of the most famous high current linac design and simulation code developed by the CEA Saclay with functions such as automatic matching and matching parameters searching, error analysis, three dimension electromagnetic field map elements, 2D and 3D particle in cell space charge force calculation routings and so on. The 3σ truncated Gaussian distribution is generated by the Tracewin code with transverse and longitudinal normalized rms emittances of 0.20 and 0.24 π.mm.mrad, respectively. The total number of macro particles is 100 000. The cavities are presented by the three dimensional field maps and the fields are calculated by Microwave Studio of CST [13]. The quadrupole lenses are presented with widely accepted hard edge ideal quadrupole lens. Three different cases with 10 mA, 30 mA and 50 mA beam currents are simulated.

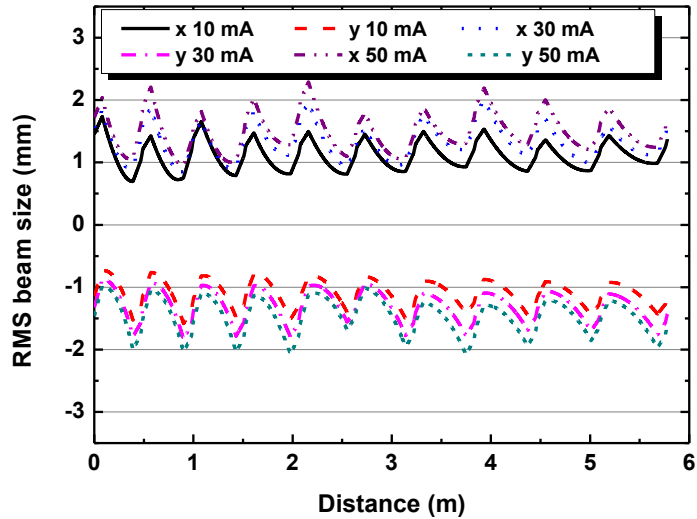

Fig.5 envelopes along the linac

Fig. 5 shows the rms envelopes of the beam along the linac. The aperture of the linac is limited by the radius of the drift tube inner radius, which is 10 mm in our case. We can see from Fig.5 that within all cavities, the rms envelopes are less than 1.5 mm, if we take 5 times rms envelopes as the total envelopes, there still enough margin for safety, even for beam current of 50 mA. Fig. 6 shows the evolution of the normalized rms emittances along the linac. We can see, for 10 mA, the emittance growths at both transverse and longitudinal direction are well controlled, and the maximum is less than 5%. As beam current increase, the emittances growth increased too. For 30 mA case, the growths are still controlled within 10%, but for 50 mA, they are increased to nearly 20%. For all three cases, the emittances growths are mainly occurred at the very beginning part of the linac, no more than 0.4 meters. That means it is occurred at the first period. This characteristic shows that they come from the charge redistribution within the beam. For the real case, the beam coming from the upstream is already reached the static distribution and the emittances growth will not happen. As comparison the normalized rms emittances evolution along the superconducting spoke012 section of the C-ADS Injector Scheme-I [14] are also shown in Fig.7. We can see, for 10 mA and 30 mA, the two linacs perform almost the same, but for 50 mA, the large longitudinal emittance growth shows it is obviously beyond the capability of the superconducting spoke012 section.

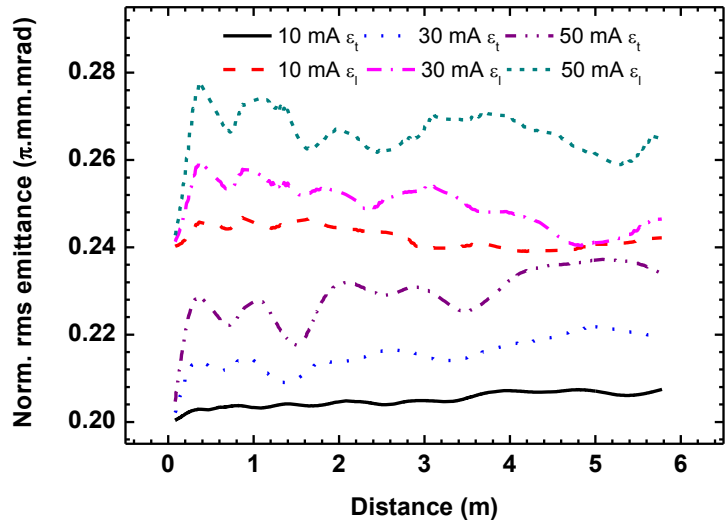

Fig.6 the normalized RMS emittances along the linac (solid: transverse; dash: longitudinal)

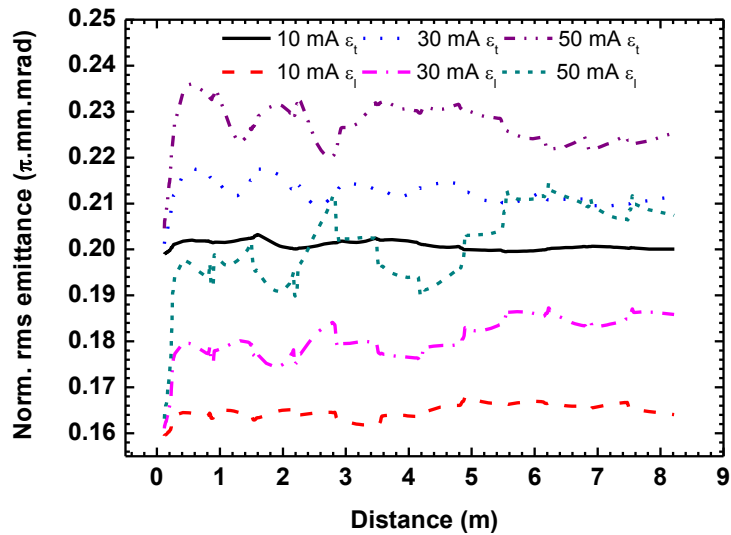

Fig.7 the normalized RMS emittances along the superconducting spoke012 section of C-ADS Injector Scheme-I (solid: transverse; dash: longitudinal)

The rms envelopes and emittances are statistic properties of the beam, they are good in describing the properties of the beam core. In order to fully describe the transport properties of the linac, Fig. 8 shows the phase distributions at the exit of the linac for three cases. From Fig.8 we can see, the transverse phase space distribution for three different beam current cases are almost identity and we can not see any distortion of the phase space distribution, except the beam halo developed as beam current increase and the halo particles can be collimated when it is necessary. On the contrary, the longitudinal phase space distributions for three different cases are quite different. As beam current increasing, the dense part of the beam core, still keeps as an ellipse, just as rms envelopes and emittances show, but at the outside, the sparse beam halos are distorted as beam current increasing. The distorted beam halo will significantly increase the total emittance. Unlike the transverse halo particles, the longitudinal ones cannot be collimated unless a dispersion section is introduced, but this will make the situation even complicated [15] and unavoidably introduce the rms emittance growth. Thus the longitudinal halo particles will transport to the high energy part. In order to keep the particle loss less than 1W/m, the halo beams have to be carefully treated and the downstream acceleration structures should have larger longitudinal acceptances. Considering the energy gain per cavity maybe larger than 10 MeV at downstream high energy superconducting sections, the beam power will be more than 500kW per cavity and the coupler for such high power is still under study. From both beam dynamics and RF technical point of view, it is better to keep the beam current less than 30 mA for the CW proton machine and the CH linac can satisfy the requirement as the front end of the CW proton linac.

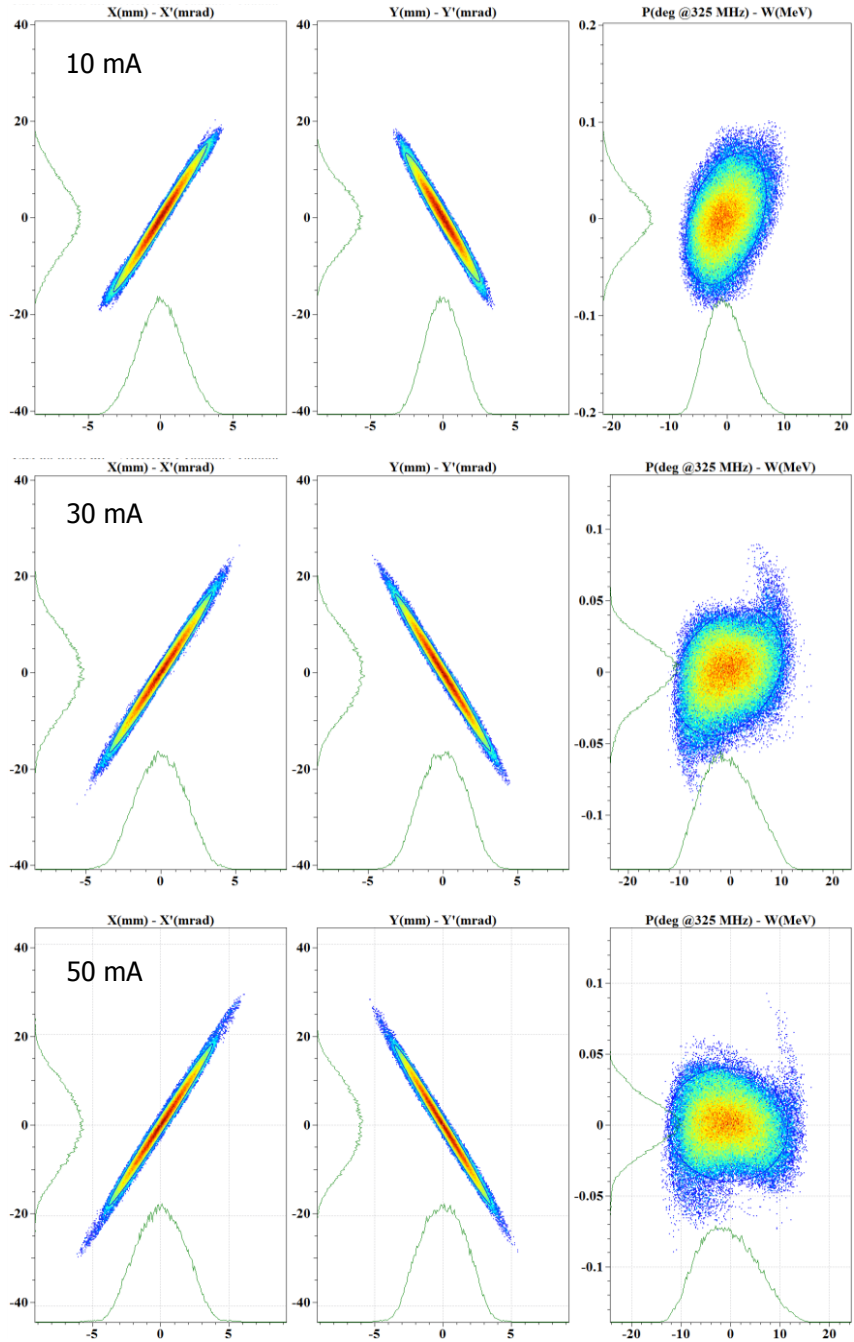

Fig.9 phase distributions at the exit of the linac with different beam current

## 5 Conclusions

The physics design for a low energy (3-10 MeV) CW proton linac based on 7-gap equidistant CH cavities is presented. The beam dynamics results are quite promising and show that it is a good candidate as front end of CW high current proton linac. By adding cavities, it is possible to accelerate up to 20 MeV with good efficiency. The only question is that if the cavity can work stably under the required gradient, and it needs to be answered by experiment. At presented, the cavity optimization and technical design of an equidistant 5-gap cavity is under way at another group in CIAE [16], and the power test is foreseeing in the future.

# 6 Acknowledgments

The author wants to thank Professor FU Shi-Nian and Professor OUYANG Huan-Fu in IHEP, Professor LI Jin-Hai in CIAE and Dr. XING Qin-Zi in Tsinghua University for their supports and discussions.

## Reference

[1] Nifenecker H et al. 2003 Accelerator Driven Sub-critical System, Institute of Physics Publishing Bristol and Philadelphia
[2] Biarrotte J.L et al. A reference accelerator scheme for ADS applications. Nucl. Instrum. Meth. A562 (2006) 656-661
[3] Carminati F et al. Report CERN-AT-93-47 (ET), CERN/LHC/96-01 (EET)
[4] Pierini, ADS Reliability Activities in Europe, OECD Nuclear Energy Agency, International workshops on Utilization and reliability of high power proton accelerators(HPPA): 4th meeting, May 16-19, Daejon, Report of Korea.
[5] Biarrotte J et al. High intensity proton SC linac using spoke cavities. Proceedings of EPAC 2002, Paris, France. 1010
[6] Lanfranco G et al. Production of 325MHz single spoke resonators at FNAL. Proceedings of PAC07, Albuquerque, New Mexico, USA. 2262
[7] Tajima T et al. Results of two LANL beta=0.175 350MHz 2gap spoke cavities. Proceedings of the 2003 Particle Accelerator Conference. 1341
[8] Garnett R et al. Conceptual design of a Low-beta SC proton Linac. Proceedings of the 2001 Particle Accelerator Conference, Chicago. 3293
[9] Zhihui Li, The focusing Properties of Both Normal and Sperconducting Low Energy CW Proton Linacs. Proceedings of SAP 2014. https://spms.kek.jp/pls/sap2014/toc.htm
[10] http://irfu.cea.fr/Sacm/logiciels/index3.php
[11] Z. Li, Design of the R.T. CH-cavity and Perspectives for a New GSI Proton Linac. Proceedings of LINAC04, Luebeck, Germany. p81
[12] Zhihui Li et al., Phys. Rev. ST Accel. Beams 16, 080101 (2013)
[13] https://www.cst.com/
[14] C. Meng et al, Error analysis and lattice improvement for the C-ADS Injector-Ⅰ, China Physics; 2014 38 (6): 067008
[15] Z. Guo, MEBT2 physics design for the C-ADS, Master Thesis, University of Chinese Academy of Sciences, 2013 (In Chinese).
[16] Jinhai LI, private communication.